# ENHANCING PAVEMENT SENSOR DATA HARVESTING FOR AI-DRIVEN TRANSPORTATION STUDIES

**Manish Kumar Krishne Gowda**
Purdue University,
465 Northwestern Avenue, West Lafayette, IN 47907
Phone: 765-767-3956
Email: mkrishne@purdue.edu

**Andrew Balmos**
Purdue University,
465 Northwestern Avenue, West Lafayette, IN 47907
Phone: 765-490-4694
Email: abalmos@purdue.edu

**Shin Boonam**
Indiana Department of Transportation,
1205 Montgomery Street, West Lafayette, IN 47906
Phone: 765-429-9225
Email: BShin@indot.IN.gov

**James V. Krogmeier**
Purdue University,
465 Northwestern Avenue, West Lafayette, IN 47907
Phone: 765-404-3120
Email: jvk@purdue.edu

*Manish Kumar, Andrew Balmos, Shin Boonam and James V. Krogmeier*

## ABSTRACT

Effective strategies for sensor data management are essential for advancing transportation research, especially in the current data-driven era, due to the advent of novel applications in artificial intelligence. This paper presents comprehensive guidelines for managing transportation sensor data, encompassing both archived static data and real-time data streams. The real-time system architecture integrates various applications with data acquisition systems (DAQ). By deploying the in-house designed, open-source "Avena" software platform alongside the NATS messaging system as a secure communication broker, reliable data exchange is ensured. While robust databases like TimescaleDB facilitate organized storage, visualization platforms like Grafana provide real-time monitoring capabilities.

In contrast, static data standards address the challenges in handling unstructured, voluminous datasets. The standards advocate for a combination of cost-effective bulk cloud storage for unprocessed sensor data and relational databases for recording summarized analyses. They highlight the role of cloud data transfer tools like FME for efficient migration of sensor data from local storages onto the cloud. Further, integration of robust visualization tools into the framework helps in deriving patterns and trends from these complex datasets.

The proposals were applied to INDOT's real-world case studies involving the I-65 and I-69 Greenfield districts. For real-time data collection, Campbell Scientific DAQ systems were used, enabling continuous generation and monitoring of sensor metrics. In the case of the archived I-69 database, summary data was compiled in Oracle, while the unprocessed data was stored in SharePoint. The results underline the effectiveness of the proposed guidelines and motivate their adoption in research projects.

*Manish Kumar, Andrew Balmos, Shin Boonam and James V. Krogmeier*

**INTRODUCTION**

Transportation research studies are aimed at exploring various aspects of network design, assessment of pavement material behavior under varying load and weather conditions (1), creating conducive environments for maintaining uninterrupted vehicular flow (2), and optimizing other key elements of traffic engineering. The primary goal of these research efforts is to improve the overall quality of transportation systems, thereby ensuring public safety. To facilitate these studies, numerous sensors are deployed across the transportation network. These sensors continuously furnish information regarding traffic dynamics and pavement conditions. Consequently, a vast amount of sensor data is generated (3-4), which needs to be effectively reduced, securely transmitted and efficiently stored. This facilitates qualitative data exploration and comprehensive interpretation using state-of-the-art data mining and machine learning techniques.

However, managing such an extensive and diverse dataset poses a major challenge since the large volume of data collected complicates the analysis (5). Furthermore, the data often necessitates intensive cleaning, validation and preprocessing to reduce errors and normalize inconsistencies that may arise due to sensor defects or varying environmental factors. Hence, effective tools and techniques to process and analyze the data are needed, preferably in real-time and in a centralized information database. Such strategies can help to better organize the data, deduce underlying patterns and relationships, employ data abstraction techniques and thereby make informed decisions based on evidence. This facilitates the development of a sustainable transportation infrastructure in the long run.

Additionally, researchers often deploy sensors from different manufacturers in the data monitoring systems. In such cases, integrating sensors with different output formats, communication protocols, power profiles, calibration specifications and interfacing mechanisms can lead to incompatibility issues. Therefore, sensor interaction, collaboration and synchronization are essential in the design of an end-to-end data management system, helping to streamline workflows and aiding proactive maintenance with minimal research costs.

Considering these constraints, in this paper, we propose overarching system architectures to address the challenges hindering seamless data monitoring and maintenance for transportation research. The long-term vision of the proposed system is to equip researchers and transportation departments with reliable and interactive tools for data management that can optimally serve their experimental and maintenance needs. We recommend guidelines to format static databases, handle real-time data ingestion and streaming, strategize sensor monitoring and data connectivity using an IoT fabric and network core, integrate agency-scale data storage solutions to organize archived and active data and incorporate interactive tools to enable comprehensive data visualization and analytics. Building on the proposed guidelines, we exemplify their utility and efficacy through two separate Indiana Department of Transportation (INDOT) projects leveraging various sensors to monitor the pavement on sections of I-65 and I-69 highways.

**MOTIVATION AND SCOPE**

Numerous transportation studies involving sensors are conducted globally each year. Many of these studies involve researchers and engineers visiting the field for experiments, physically monitoring the sensors for extended periods and then manually transporting the collected data back to a central location for analysis. This approach is inherently burdened with multiple drawbacks. In addition to the labor-intensive process of physically overseeing the experiments and offloading the data for post-processing, researchers often experience data loss or erroneous data. Deferred processing and delayed experimental feedback, risk overlooking the defective data that may arise from factors such as hardware failures, power interruptions, environmental elements or even human errors. These data irregularities can be prevented by automated real-time monitoring at the central data storage location. Proactive monitoring facilitates early detection of outages, sensor malfunctions and local storage anomalies. Further, real-time data visualization tools provide immediate pavement health insights and deeper situational awareness, minimizing post-processing efforts. This motivates the need for a comprehensive system for real-time data monitoring and delivery.

A wealth of sensor data is available in various storage repositories throughout the industry, primarily in a non-standard, unprocessed state. We refer to such pre-compiled, historical data logs as static datasets throughout this literature. Researchers and engineers would often require viewing the summary of these static sensor databases, including critical details, such as timestamp, data quality indicators, calibration information, key error logs and other sensor metadata. Accessing, navigating, and manipulating existing sensor records is fraught with significant challenges, given the enormity of size and irregularities in data archiving. An industry standard to manage large-scale, unprocessed sensor databases would greatly enhance access to the data and prepare a data record suitable for modern processing techniques incorporating artificial intelligence and machine learning. Such guidelines would not only help researchers to better organize the sensor data and identify critical records, but also help to identify key patterns and trends, assess the data quality and make informed decisions. Further, given the monumental scope of the data, software tools to efficiently transfer the data to and from the database is imperative. This will improve the responsiveness of the system and expedite the research process. Similar to a real-time data system, integrating reliable data visualization tools with such static databases will help in interactive data analysis and comparative research.

Consolidating the static database management and real-time data collection and monitoring process under industry-wide standards will help not only the researchers, but also transportation planners who are interested in big data analysis. Such organizational standards must not only be robust, proposing widely accessible and adaptable tools, but also be flexible and futuristic, composed of state-of-the-art, preferably open-source applications. It should also be compatible with existing solutions, ensuring seamless integration and simple upgradation of the prevalent techniques. This will enhance capabilities, promote scalabilities, improve performance and efficiency and further bolster innovation, maximizing return on investment in the long run. Enabling the researchers to take a data-driven approach will further optimize resource allocation and support predictive analysis with artificial intelligence and expand the scope of transportation research.

## METHODOLOGY

In this section we first propose the operational standards for static database management and the system architecture for managing real-time data. We further illustrate the application of these guidelines through practical deployments in two INDOT projects.

### Guidelines for Static Data

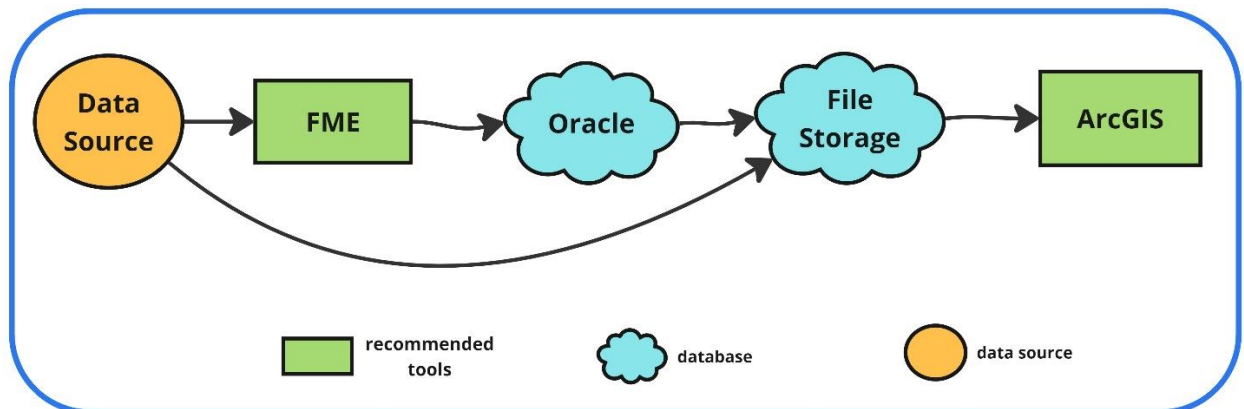

**Figure 1 Proposed guidelines for static database management**

Figure 1 describes generic guidelines for managing unorganized databases arising from intensive research efforts that have the potential for providing valuable information regarding the transportation landscape upon comprehensive analysis. The guidelines outline best practices and strategies for handling such static databases. Following are the key components of the proposal:

1. **Data source:** Unlike the real-time data source, a static data source is of known size and fixed capacity, with the sensor sampling parameters, data acquisition conditions and measurement settings either chronicled and archived within the database or in many cases, poorly documented. Considering the extensive volume of data, it is often preserved and maintained by researchers in its preliminary, unscrutinized form in tangible storage media such as hard drives and storage disks or primitive online repositories. Local storage not only limits data accessibility and collaboration capability, but also raises the risk of data loss due to hardware failures, thereby increasing the cost of research in the long term. Desktop data processing tools such as Excel, Matlab, Python or Apache Hadoop are used for static data management and analysis, often with minimal online support. The data analysis potential is restricted by the limited computational resources and processing powers of personal workstations. Bringing these static data sources under the ambit of a robust online framework can greatly benefit from a multitude of web-based data analytics tools that can potentially integrate with the cloud storage solutions and simultaneously reduce the security risks associated with in-house data management.

2. **Cloud data transfer tools:** As briefly outlined in the motivation section, employing efficient and reliable data transfer tools is essential for sensor data migration into the online realm. In addition to ensuring timely data transfer while maintaining data fidelity during the process, the transfer tool must be compatible with data files of varying sizes and formats. Flexibility of automated scheduling would enable the researchers to manage bulk transfers with minimal manual intervention, refraining from overloading the local resources or network bandwidth. Error detection and logging features, addressing any network complications during the data transfers, safeguards against data corruption or degradation. While scripting languages like Python facilitate customized and efficient data transfers aided by wide-ranging API compatibility, the steeper learning curves resulting from the necessity to master language specific syntax and semantics along with complex data transfer logics would limit the adaptability of the framework by research communities with limited programming expertise. Further, it could delay knowledge exchange and skill transfer between researchers, impacting organizational learning and development. Therefore, integrating tools with user-friendly and intuitive graphical interfaces that require minimal or no programming skills, while ensuring the manifestation of the key data transfer features discussed thus far, is essential. Syncthing, FME, CloudHQ and CloudFuze are some of the popular solutions that support data migration and transfer management.

3. **Databases:** With a large amount of sensor data available locally at the researchers' disposal, hosting the data on cloud storage significantly enhances the data processing and resource management capabilities, as already discussed. In order to assist a holistic, broad-scope analysis of the extensive dataset, it would be beneficial for the researchers to form a rudimentary understanding of the breadth and scope of the data under consideration. Such insights on the executive synopsis of the sensor dataset will help the researchers to better plan and strategize a comprehensive and in-depth analysis of the research data. Further, a concise overview of the telemetry data can help in data quality assessment, preliminary error detection, identification of gaps and irregularities and formulation of hypotheses, thus facilitating resource optimization. Analytical summary of datasets, including temporal data patterns, sensor metadata, performance metrics and descriptive statistics are frequently accessed by the researchers throughout the course of their study. To respond to this need, we propose leveraging cost-effective bulk cloud storage solutions to cater to the larger unprocessed sensor dataset and relational databases for hosting summarative analysis of the data (12). Relational databases stores data in tables in an organized and systematic method, featuring options to maintain logical interlinks between correlated pertinent data. These databases, although expensive, provide robust features for database wide querying and managing the formatted data using a structured query language (SQL). This simplifies the process of reviewing key sensor parameters and analyzing data behavioral patterns and predictive trends. MySQL, Oracle and PostgreSQL are some of the popular relational databases available for research use. Bulk or blob storage solutions on the other hand support large-scale, multi-format data storage with built-in sophisticated security features and data recovery options. They provide flexible pricing options and adaptable storage tiers, making it suitable for transportation sensor research. Data can be easily interlinked with relational databases, providing a holistic storage solution for static sensor databases. SharePoint, Purdue RCAC data depot, AWS, Google cloud, etc. are some of the blob storage solutions explored for formulating the static framework.

4. **Visualization tools:** The generic features of visualization tools necessary for modeling static sensor data are similar to real-time data visualization tools. However, the nature of sensor data can dictate the choice of tools for conducting visual analysis in the context of static databases. For example, data with high granularity can demand tools supporting smoothing functions or interpolation methods to reduce noise, while for extremely large datasets, the visualization tools must handle scale effectively using techniques like down sampling or interactive filtering. It is essential to account for both retrospective and prospective data points when providing chronological and futuristic insights. On the other hand, spatial data arising from sensors such as GPS, cameras and inertial measurement units require tools such as ArcGIS or QGIS, capable of discerning geographic trends and regional dynamics for making location-sensitive decisions.

*Demonstrative case study*

The proposed static dataset guidelines were applied to administer two databases maintained by separate Purdue-INDOT joint research teams. The first database consists of pavement sensor data from field trials on sections of I-69 highway while the second contains historical logs from accelerated pavement testing (APT) at the INDOT division of research, Lafayette.

a) The I-69 database contains pressure cell and strain gauge data from multiple experiments, stored in a MATLAB compatible (.mat) file format. Each experiment involved embedding the sensors in different sections of the pavement and obtaining their response to varying controlled loads for a restricted amount of time. The data was collected by researchers over a period of three years, from 2020-2022. Researchers had developed a MATLAB application to examine and visualize the data. The program denoises the data, identifies the maxima and minima values for each standalone experiment. A sample graph of the data from an experiment is shown in Figure 2. Due to the extensive size of the data and limited computational power of the desktop, the application would frequently become unresponsive. Further, performing comparative analysis across different experiments using the application was challenging.

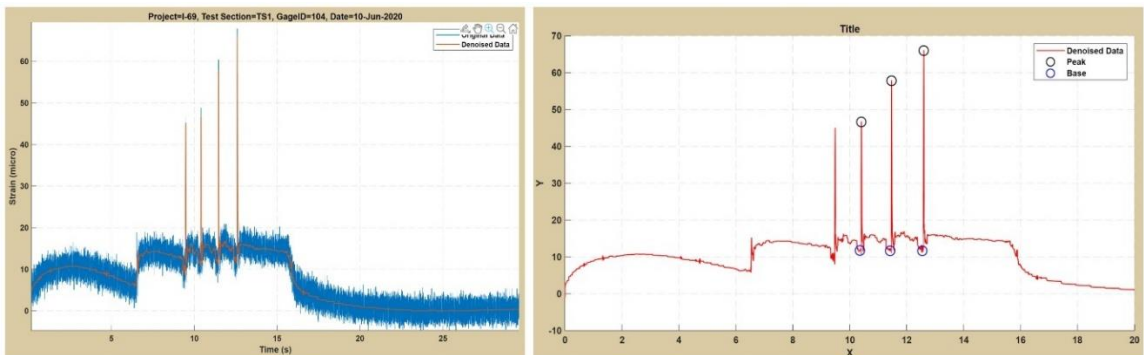

**Figure 2 Sample view of sensor data alongside denoised data (left) and captured peaks and troughs (right)**

To address these limitations, the coordinates of the peaks and troughs of the pressure cell and strain gauge were captured in a single tabular format. Additionally, key data such as file name, test section ID, sensor ID, sensor type, location information, date of data capture, load size, calibration information, etc. that provide valuable insights about the sensor data and could aid further analysis were also captured alongside in the structured table format. This format consolidates relevant information and summarizes the database, enabling the researchers to perform cross-analysis of sensor data across different files. A sample view of the processed tabular data is depicted in Figure 3.

The processed tabular data and the unprocessed MATLAB files were stored in Microsoft SharePoint, demonstrating an instance of blob storage explained earlier. This ensured a centralized and readily accessible storage for the entire research team.

| Filename | Captured instance | Asphalt Strain Gage | Placement | Cal Coeff(x10-6/1x10-6) | Rated Output(x10-6) | Extrema | SecondsElapsed | processed_datapoint(microstrain) |
|---|---|---|---|---|---|---|---|---|
| Traffic D1 F20 07-07-22.txt | first20 | 7 | 36 | 0.849 | 5890 | maxima | 9.7004 | -0.185732682 |
| Traffic D1 F20 07-07-22.txt | first20 | 7 | 36 | 0.849 | 5890 | maxima | 19.5662 | -0.184823931 |
| Traffic D1 F20 07-07-22.txt | first20 | 7 | 36 | 0.849 | 5890 | maxima | 30.9986 | -0.183044011 |
| Traffic D1 F20 07-07-22.txt | first20 | 7 | 36 | 0.849 | 5890 | maxima | 42.6466 | -0.18189681 |

**Figure 3 Sample view of the formatted pressure cell and strain gauge data**

b)  The second database containing APT experiment data, was significantly larger and more comprehensive. This database contains readings from different categories of sensors, collected from INDOT's APT section. The sensor dataset can help the researchers to study pavement performance. Following is a brief overview of the different sensor categories of the dataset and the formatting approach undertaken to align the data with the proposed static data guidelines.

i.  Asphalt Laser Profile Data: This comprises of laser readings sampled from different sections of the asphalt segment of the pavement. To minimize the footprint, only the first 20 passes and the last 20 passes of 1000 passes of the prototype wheel loading were captured, amounting to 4000 samples per experimental recording. The readings aid researchers in their study of vertical profile of the asphalt pavement surface. A custom algorithm was designed to filter the collected data. Specifically, a Savitzky-Golay filter with a window size of 1000 was employed to smoothen the laser profile data. The window size was determined through trial and error to ensure effective noise reduction without significant data distortion.

After applying the Savitzky-Golay filter, all the samples along with their corresponding sampled time instants were retained, as they are essential for realizing the general horizontal profile of the pavement surface. A sample graph of the Asphalt Laser profile data is shown in Figure 4 and the formatted table is shown in Figure 5.

ii.  Asphalt Pre-Traffic Laser Profile Data: This data is similar to the Asphalt Laser Profile Data. It contains laser readings which are taken before initiating the 1000 passes. By capturing pre-traffic data, researchers can assess the pavement's preliminary condition and anticipate changes that could occur during the testing process. Graphs and table formats are similar to Figure 4 and Figure 5 respectively.

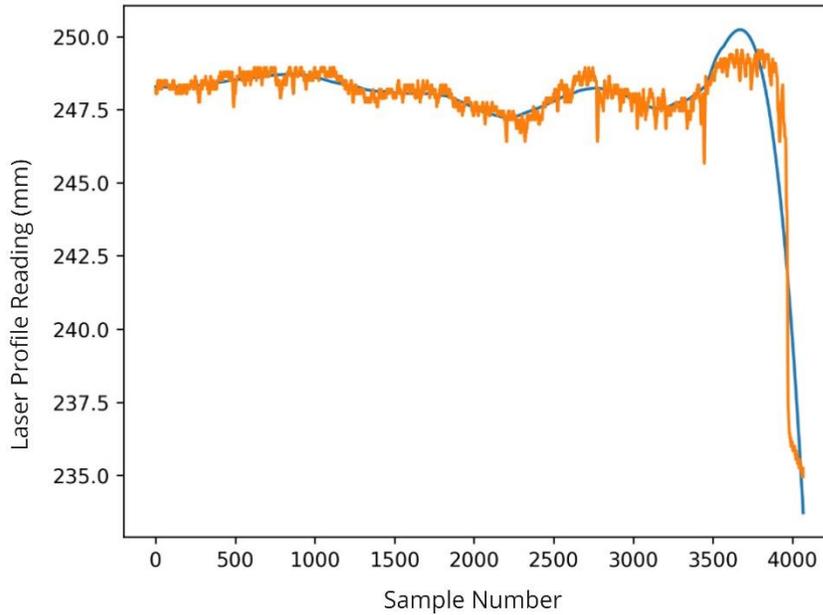

**Figure 4 Unprocessed (orange) and filtered (blue) data of an Asphalt Laser Profile experiment**

| Filename | Sample Number | Horiz(mm)=samp_no*(1384/8088) | Laser reading(mm) | Laserbeam locn(mm) | Sampled Time |
|---|---|---|---|---|---|
| 19-06-18 H L1 0-Passes PL1 - 1400mm 1 10-57.xls | 1 | 0.171117705 | 250.8819079 | 0 | 10:57:16.47 |
| 19-06-18 H L1 0-Passes PL1 - 1400mm 1 10-57.xls | 2 | 0.34223541 | 250.8802211 | 20 | 10:57:16.50 |
| 19-06-18 H L1 0-Passes PL1 - 1400mm 1 10-57.xls | 3 | 0.513353116 | 250.8785381 | 20 | 10:57:16.52 |
| 19-06-18 H L1 0-Passes PL1 - 1400mm 1 10-57.xls | 4 | 0.684470821 | 250.8768587 | 20 | 10:57:16.55 |

**Figure 5 Formatted table of Asphalt Laser Profile**

iii. Asphalt Sensor Data: This dataset includes readings from various sensors installed in the asphalt pavement. It captures essential information about its structural behavior and response to loadings. The database contains measurements from four types of sensors, viz. Asphalt Strain Gauge (ASG), Concrete Strain Gauge (CSG), Pressure Cell (PC), and Thermocouple (TC). These sensors are strategically placed beneath the pavement, along different longitudinal and vertical sections and the data is recorded for the first 20 passes and last 20 passes of the load. To facilitate analysis, four different tables corresponding to each sensor type are created.

The collected data is accompanied by units of measurement for each sensor type, viz. microstrain for ASG, inches for CSG, KPa for PC, and Fahrenheit for the Thermocouple. The Savitzky-Golay filter is applied with a window size of 1000 for smoothening the ASG and CSG sensor data. However, smaller window sizes of 50 and 100 are used for the Thermocouple and PC sensors, respectively.

For each unprocessed file in the local database, 20 peaks from the first 20 passes and 20 peaks from the last 20 passes are identified for ASG, amounting to 40 peaks. Rather than the bases, which were relevant in the I-69 pressure cell and strain gauge database, the elastic recovery region is needed for analysis of the pavement behavior. In the context of pavement analysis, the elastic recovery region refers to the resilient behavior of the pavement surface after the application and subsequent removal of the load. The elastic recovery region is captured by identifying specific points on the response curve of the pavement data. These points represent the envelope of the elastic recovery, which is a more relevant entity for analysis compared to the minima of the response curve. Thus, to capture the envelope of the elastic recovery, multiple points (approximately 5 points for each pass) corresponding to these envelopes are recorded. The resulting 40 peaks and the elastic recovery points (approximately 200 for each file) were saved in the formatted file. For CSG, PC and Thermostat, only maxima and minima are captured.

A sample graph of the asphalt sensor is shown in Figure 6 while the formatted table is shown in Figure 7.

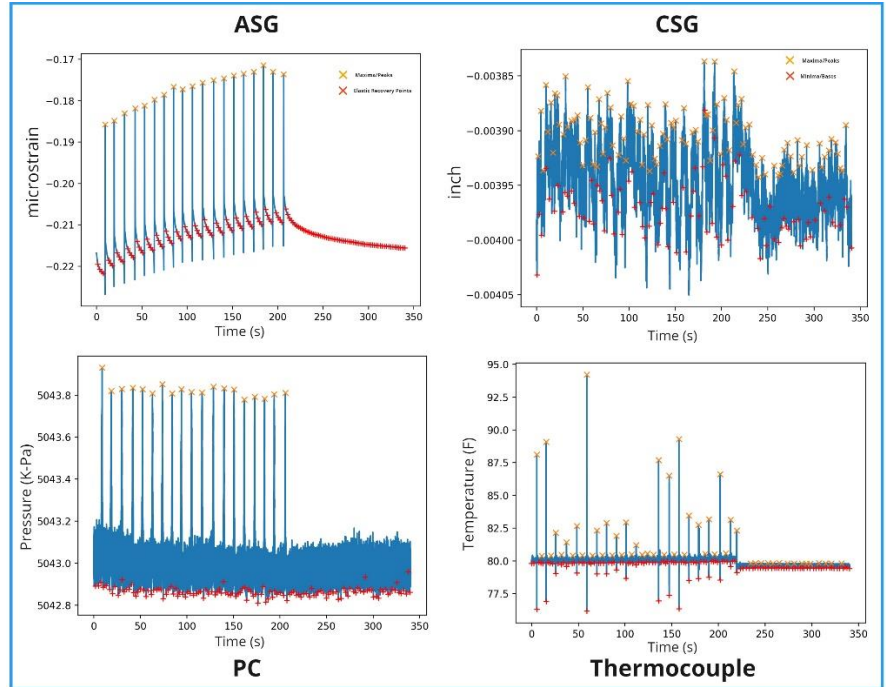

**Figure 6 Sample view of maximas and minimas of Asphalt Sensor Data**

| Filename | Captured instance | Asphalt Strain Gage | Placement | Cal Coeff(x10-6/1x10-6) | Rated Output(x10-6) | Extrema | SecondsElapsed | processed_datapoint(microstrain) |
|---|---|---|---|---|---|---|---|---|
| Traffic D1 F20 07-07-22.txt | first20 | 7 | 36 | 0.849 | 5890 | maxima | 9.7004 | -0.185732682 |
| Traffic D1 F20 07-07-22.txt | first20 | 7 | 36 | 0.849 | 5890 | maxima | 19.5662 | -0.184823931 |
| Traffic D1 F20 07-07-22.txt | first20 | 7 | 36 | 0.849 | 5890 | maxima | 30.9986 | -0.183044011 |
| Traffic D1 F20 07-07-22.txt | first20 | 7 | 36 | 0.849 | 5890 | maxima | 42.6466 | -0.18189681 |

**Figure 7 A view of the formatted table of the Asphalt Sensor Data**

iv.  Asphalt Stationary Load Data: This section provides data related to stationary load tests applied to the asphalt pavement. It offers insights into its bearing capacity and deformation characteristics. Two types of loads are used for impact testing: end tire (ET) and mid tire (MT). ET simulates the impact on the sensors when they are directly beneath one of the tires, while MT simulates the impact when the sensor is positioned between the axles of the two tires. The processing of maxima and minima for the asphalt stationary load data is similar to that of the asphalt sensor data. Peaks are identified, and envelope points are captured for both ET and MT scenarios.

A sample view of the formatted table for an ET load is shown in Figure 8.

| Filename | Asphalt Strain Gage | Placement | Cal Coeff(x10-6/1x10-6) | Rated Output(x10-6) | Extrema | SecondsElapsed | processed_datapoint(microstrain) |
|---|---|---|---|---|---|---|---|
| Stationary Load Pt 0 ET.txt | 7 | 36 | 0.85 | 5880 | maxima | 10.28 | -0.412719961 |
| Stationary Load Pt 0 ET.txt | 7 | 36 | 0.85 | 5880 | maxima | 20.281 | -0.41226282 |
| Stationary Load Pt 0 ET.txt | 7 | 36 | 0.85 | 5880 | maxima | 30.295 | -0.411937625 |
| Stationary Load Pt 0 ET.txt | 7 | 36 | 0.85 | 5880 | maxima | 40.304 | -0.411608014 |

**Figure 8 A view of the formatted table of the Asphalt Stationary Load Data**

v. Concrete Laser Profile Data: This section consists of laser profile data for concrete segment of the pavement. These readings help to assess the vertical profile of the concrete surface. The data formatting is similar to the asphalt laser profile data:.

vi. Concrete Pre-Traffic Laser Profile Data: This section captures pre-traffic readings for the concrete pavement. The data formatting is similar to the asphalt pre-traffic laser profile data.

vii. Falling Weight Deflectometer (FWD) Related Sensor Data: This dataset comprises readings from sensors associated with the FWD equipment used for non-destructive pavement evaluation. The FWD data replaces the moving load with the FWD apparatus.

A sample graph of the FWD data is shown in Figure 9 while the formatted table is shown in Figure 10. Data formatting for the APT database was performed using Python language.

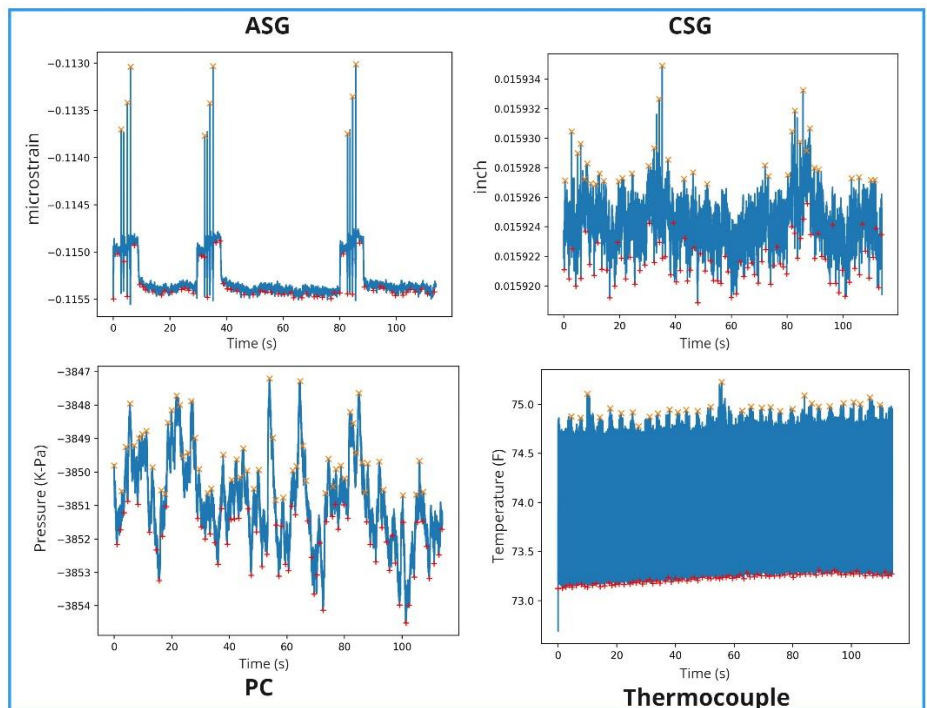

**Figure 9 Sample view of maximas and minimas of FWD Data**

| Filename | Asphalt Strain Gage | Placement | Cal Coeff(x10-6/1x10-6) | Rated Output(x10-6) | Extrema | SecondsElapsed | processed_datapoint(microstrain) |
|---|---|---|---|---|---|---|---|
| FWD Pass 1 #0 06-10 TRY 1.txt | 6 | 35 | 0.847 | 5900 | maxima | 2.8349 | -0.113703747 |
| FWD Pass 1 #0 06-10 TRY 1.txt | 6 | 35 | 0.847 | 5900 | maxima | 4.9983 | -0.113419536 |
| FWD Pass 1 #0 06-10 TRY 1.txt | 6 | 35 | 0.847 | 5900 | maxima | 6.0862 | -0.11303932 |
| FWD Pass 1 #0 06-10 TRY 1.txt | 6 | 35 | 0.847 | 5900 | maxima | 32.3699 | -0.113770396 |

**Figure 10 A view of the formatted table of the FWD Data**

Oracle database, representing an instance of the relational database component of the static data guidelines, was used to store the formatted tabular data, utilizing Safe's FME software as the data transfer tool.

Oracle tables were created using the SQL "CREATE TABLE" command. During the process of these tables' creation, careful consideration was given to the data types assigned to each field. Without relying on default data types, specific data types were selected intentionally to match the data being processed. This approach offers greater control over each field's size, subsequently providing a clear understanding of the overall table size. Unlike dynamic sizing based on varying fields, this method ensures predictability and efficiency in terms of memory allocation.

An important aspect of the data storage strategy in the relational database is the application of normalization techniques. In practice, information related to file details is stored within a separate table. This separation helps conserve storage space, as the file information doesn't need to be duplicated for every data point in the main table. Instead of repeatedly storing redundant file details, a unique identifier (referred to as the "filename_id") is assigned to each data point. This identifier acts as a foreign key, establishing a link between the individual data points and the table furnishing the file information.

This approach has multiple benefits. First, it minimizes data redundancy and optimizes storage efficiency. Second, querying the database to retrieve filename-related information, utilizing the "filename_id" renders the process straightforward and efficient. The normalization technique expedites the retrieval of associated file information. A view of the created Oracle table is shown in Figure 11.

**Figure 11 Sample view of an Oracle table containing formatted data (left) and depiction of database normalization (right)**

The FME software developed by Safe is employed for the upload of processed data and file information into an Oracle database. FME is an efficient tool for data transformation, migration and integration tasks.

The process begins with the software reading the CSV file containing the summarized data. FME uses a dedicated module called a "reader" to efficiently parse the contents of the CSV file. As the processed file is parsed, unique filenames are identified and logged.

The unique filenames that are logged are extracted and subsequently uploaded to a designated table named "FILE_INFO" in the Oracle database. This table serves as a repository for the file-related information, avoiding the need to store redundant information for each data point, as described earlier. The file information may include attributes such as file name, creation date, author, and any other relevant sensor metadata.

To link the processed data with their respective file information, an essential module in FME called the "database joiner" is employed. This module utilizes the unique filenames as a reference and queries the "FILE_INFO" table in the Oracle database. By comparing the filenames, the module retrieves the corresponding IDs assigned to each unique file. These IDs act as foreign keys that establish connections between the processed data and their associated file information.

The data upload process is designed for efficiency, benefiting from FME's optimized modules and workflows. The software offers a wealth of additional modules that can be further explored to enhance the data integration process. These modules are purpose-built to handle various data manipulation tasks, allowing for customization and optimization of workflows according to specific requirements. Furthermore, the graphical user interface facilitates an intuitive understanding of the data transfer process.

A visual representation of the data upload process is provided through Figure 12 illustrating the FME workbench. This workbench showcases the sequence of modules, connections, and data flow involved in the upload process. The figure offers a clear visualization of how data transformation and integration are executed within the FME environment.

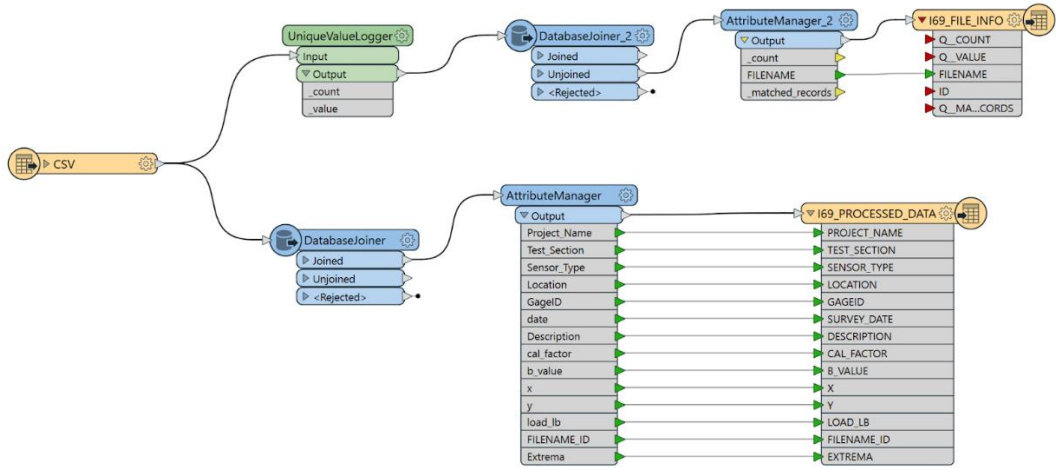

**Figure 12 FME workbench for I69 database**

ArcGIS, a widely adopted tool among INDOT researchers, can be utilized to visualize the data stored within the Oracle database. ArcGIS serves as a platform that facilitates the representation of complex geospatial data through visualizations. A sample ArcGIS project, currently in the inceptive phase, developed using the formatted data is shown in Figure 13.

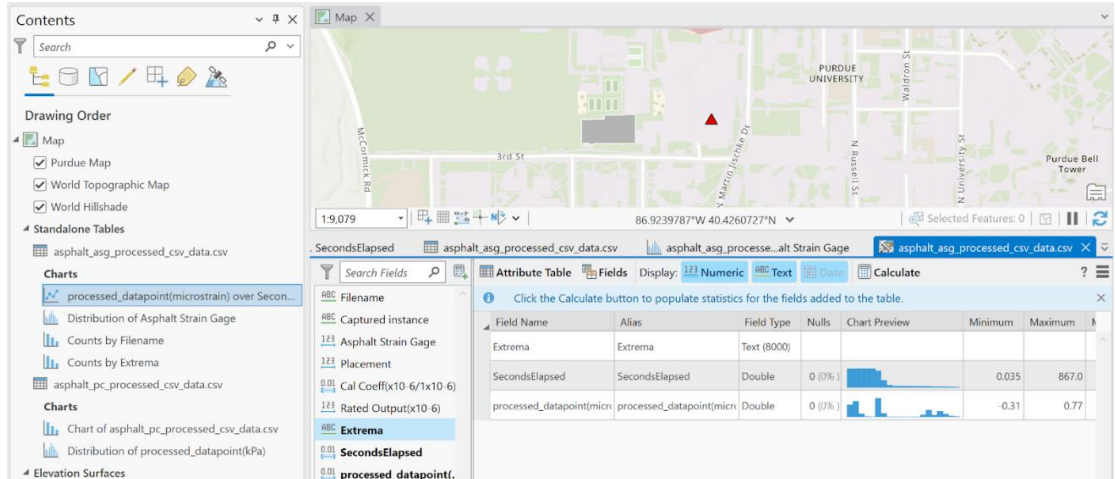

**Figure 13 Basic visualization of processed data in ArcGIS**

A system workflow for the static data framework, showing the interconnection of components along with the deployed tools is presented in Figure 14.

*Manish Kumar, Andrew Balmos, Shin Boonam and James V. Krogmeier*

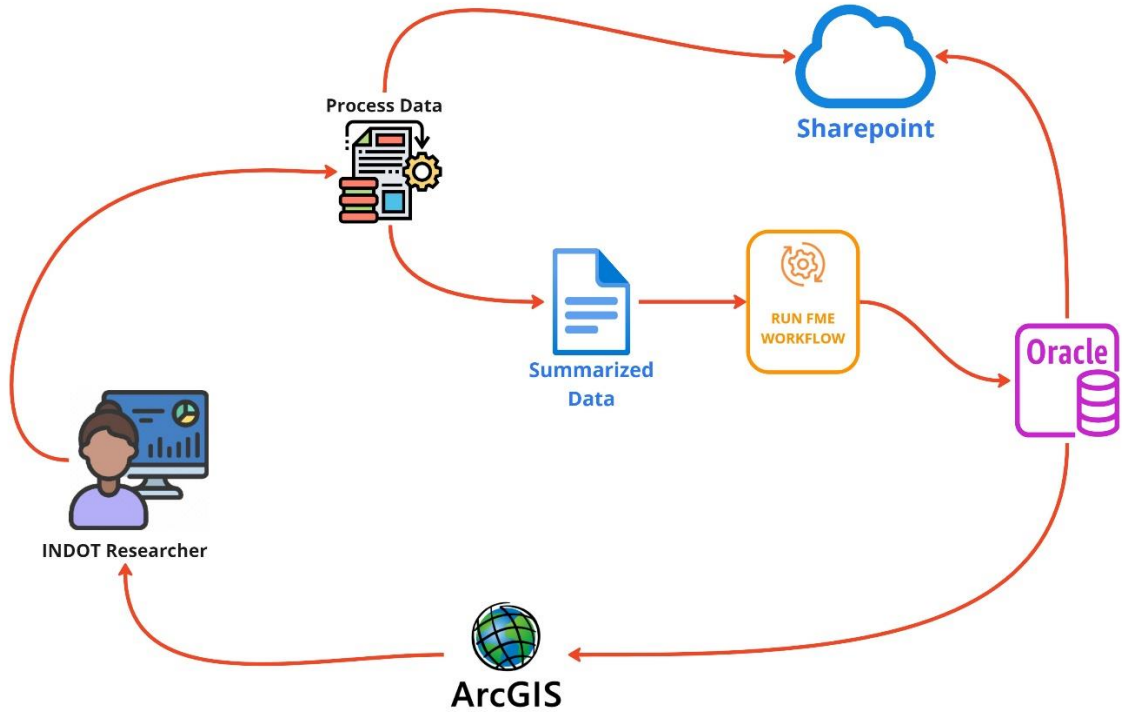

**Figure 14 Sample workflow for static data processing**

**System Architecture for Real-time Data**

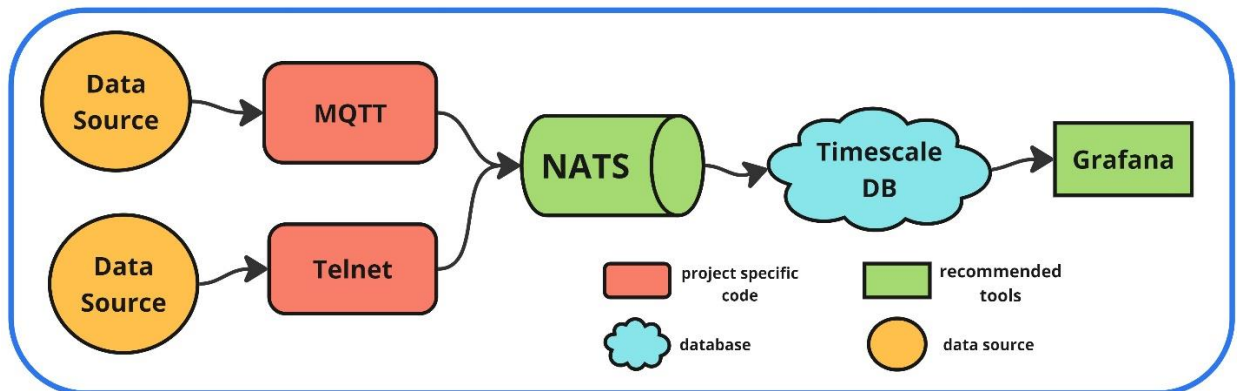

**Figure 15 Proposed system for real-time data management**

While managing static databases necessitates dedicated preprocessing and management, real-time sensor data collection from experimental sites presents unique challenges as well. Figure 15 describes a comprehensive end-to-end solution for optimizing the remote monitoring of the real-time sensor data generated at the data source. The proposed system leverages "Avena", a novel open-source framework that can intelligently process high-resolution data (8-9). Avena was initially designed to create an efficient edge computing ecosystem for machine computers, while safeguarding data autonomy. The framework operates on Linux containers and cloud technology and facilitates data flow between edge computers and server systems, ensuring seamless data exchange in the edge-to-core architecture. Although Avena was initially envisioned for agricultural platforms, it can be seamlessly adopted for data management in transportation research systems.

The proposed architecture for real-time data collection and management is proposed to be generic, comprising of multiple adaptable components that can be customized for different operational instances. The Avena framework describes how data should flow between devices and presents an overlay network in which all the components on the network appear to be directly connected, regardless of their potential disconnection or spatial distribution. Following is a brief of these constituents of the architecture:

1. **Data source:** This refers to the system generating sensor data enabling analysis and research. A multitude of sensors are deployed for transportation studies, including earth pressure cells (EPCs), strain gauges, video cameras, humidity sensors, air quality sensors and laser profile sensors. Combinations of sensors are selected based on the specific research problem that needs to be addressed. A data acquisition system (DAQ) (7) actively monitors these sensors. The DAQ consists of signal conditioning units, analog-to-digital converters and power management modules and is capable of sampling and collecting the sensor data. Different vendors such as Labjack, Campbell Scientific, BDI and National Instruments provide commercial DAQ solutions.

2. **Data interface:** This pertains to a software application hosted on the DAQ that facilitates the transmission or exchange of sensor data between the DAQ and the NATS messaging system. Depending on the sophistication, it can represent a standard messaging protocol such as message queuing telemetry transport (MQTT) or Telnet. NATS is a secure, lightweight messaging service designed for creating multi-node distributed platforms, cloud-native applications and IoT solutions (6). NATS is widely integrated into IoT systems due to its scalability and robust performance. Further, being an open-source messaging system supporting multiple programming languages, it offers greater flexibility for researchers and developers. These considerations establish NATS as an optimal choice for integration into the primary pathway for service interactions in the real-time system architecture. Here, NATS operates as a message broker in publish-subscribe model, whereby the DAQ communicates sensor messages to a particular topic and the server subscribing to the same topic receives the sent messages, thereby avoiding any redundant data exchanges in the system.

3. **Database**: Integrating a robust database within the end-to-end system provides an organized and efficient storage solution and allows for convenient access of the data by the researchers in a secure, readily accessible repository. Ideally, a potential candidate for a reliable database should be capable of managing high-speed time-critical sensor data, accommodate increasing number of sensors, adapt to larger datasets, handle queries at scale and support notification and alert features. Several contemporary database solutions enable these benefits and additionally facilitate real-time coordination with data analytics and visualization tools, artificial intelligence models and third-party APIs, all of which can significantly bolster the data analysis potential of researchers. Notable examples are InfluxDB, TimescaleDB, Apache Pinot and Prometheus.

4. **Visualization platform:** Data representation and visualization tools that seamlessly integrate with the adopted database are essential for real-time monitoring of the sensor measurements. These tools enable researchers to conduct preliminary assessment of the sensor data and evaluate if the patterns and trends are consistent with the expected behavior. Distinctive features such as user-friendly interface, interactive and customizable dashboards, versatile visualization options including graphs and maps, high frequency data processing competence, data filtering and aggregation capabilities, cross platform compatibility are some of the valuable prospects that are desirable in these tools. Power BI, Grafana, Kibana and Tableau are some of the popular visualization platforms used for data illustration and graphical representation of data streams.

*Demonstrative case study*

The system architecture was integrated into the sensor data acquisition trials, overseen by INDOT and Purdue researchers in two experimental sites on I-65 highway demonstrating the functional aspects of the conceptual design. Soil compression gauges (SCG) from Geocomp, earth pressure cells (EPC) by Geokon and TEROS-12 soil moisture and temperature sensor were embedded in two separate sections of I-65 pavement. These sensors, among other aspects, are used to assess the traffic load distribution on the pavement, monitor structural stability of construction materials, evaluate the response of pavement components to shifting weather and climatic conditions, study soil deformation characteristics under various loading scenarios and predict pavement failures ahead of time. Figure 16 and Figure 17 depict a sample placement of the sensors in the I-65 experimentation section for the research study.

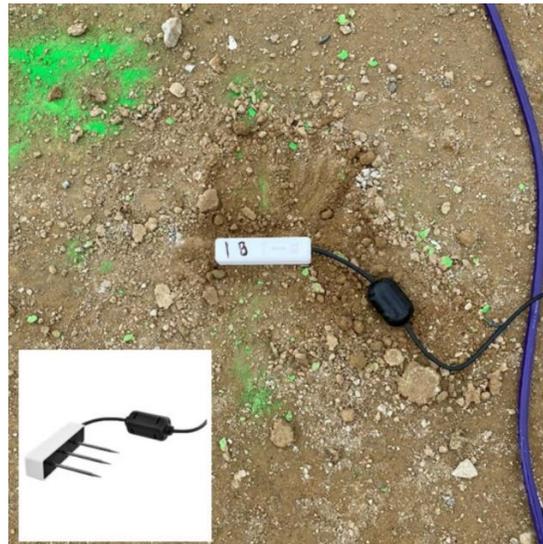

**Figure 16 Teros-12 soil moisture and temperature sensor**

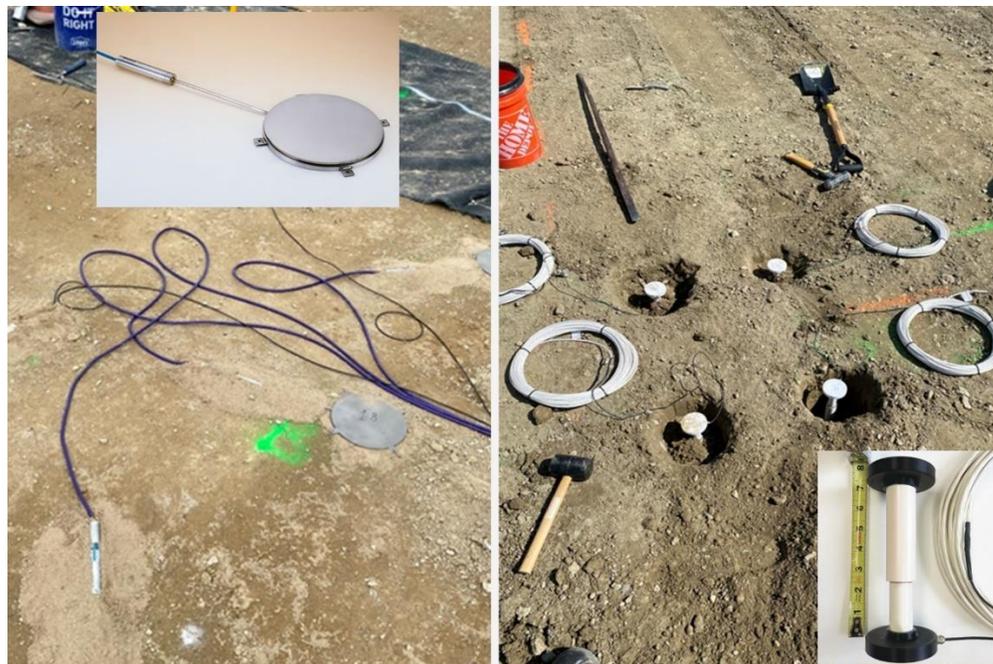

**Figure 17 Geokon GK4810-1 vibrating EPC (left) and Geocomp GC-SCG-13 SCS (right)**

A Campbell Scientific DAQ was installed on the roadside to sample and monitor the embedded sensors. The DAQ features a cell modem that provides internet connectivity, facilitating wireless data transmission to remote servers. It also accompanies a solar panel to power the sensors and the internal components of the DAQ and a data logger for sensor data acquisition and storage. The Campbell Scientific DAQ used for the real-time data management architecture demonstration is shown in Figure 18.

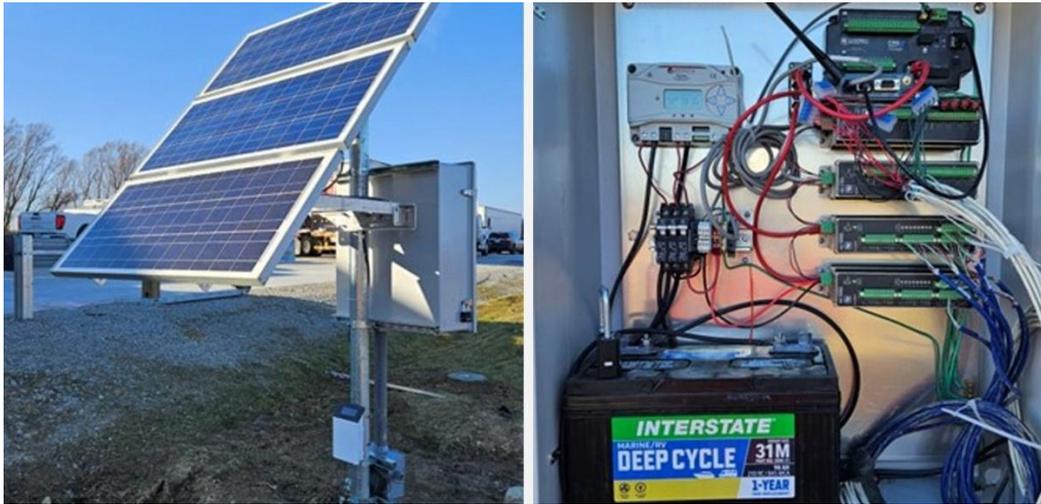

**Figure 18 Campbell Scientific DAQ with the solar panel (left) and CR6 data logger (right)**

The Campbell Scientific DAQ system offers built-in support for MQTT data transmission. MQTT is a TCP/IP based messaging protocol, primarily used in IoT applications to exchange data between interconnected sensors and server systems. NATS, being compatible with MQTT protocol, enabled seamless interconnection of the output sensor data stream from the DAQ with the database. In general, the flow consists of the Campbell Scientific DAQ connecting via MQTT to a centralized hosted NATS server. Due to a limitation of the DAQ hardware, only one-second time averages of sensor readings can be published to sensor specific MQTT topics, and, therefore, NATS subjects. An open-source software application, RedPanda Connect, running continuously at the centralized location, receives the data payloads via wildcard NATS subscription on all sensor subjects. The data format is slightly transformed and inserted into a TimescaleDB database. For this demonstration, Grafana queries the TimescaleDB database for both real-time monitoring and visualization purposes. In this manner, NATS decouples sensor type, location, data transmission and data storage, enabling each component to operate and scale independently.

TimescaleDB is an open-source database, developed on top of PostgreSQL and is optimized to manage large scale time-series data (10). It organizes the incoming data stream efficiently by partitioning the data both spatially and temporally. This feature renders it an ideal storage solution, particularly given the need to process large volumes of high-speed data from a specific set of sensors and support complex analytical queries. Some of its other notable features include support for data compression, configuring alerts based on specific thresholds, compatibility with data visualization tools like Grafana and Tableau and the ability to perform predictive analysis with the data. Grafana, selected as the visualization tool (11), is a versatile, open-source, web-based visual analytics software, supporting the creation of customizable dashboards, interactive graphs and informative tables on the time-series data. It features a user-friendly drag-and-drop interface with a range of pre-packaged extensions and custom panels to assist layout design. It also allows researchers to share the dashboards, along with options for role-based access, facilitating collaborative supervision and broad-scale inspection of the sensor data stream. A sample view of a Grafana dashboard, graphically depicting the real time sensor data flow is shown in Figure 19. More than 2 million sensor samples were recorded in the database, per day, per field trial location. A nominal sub second display latency was observed between the data sampling instance and its display on the dashboard.

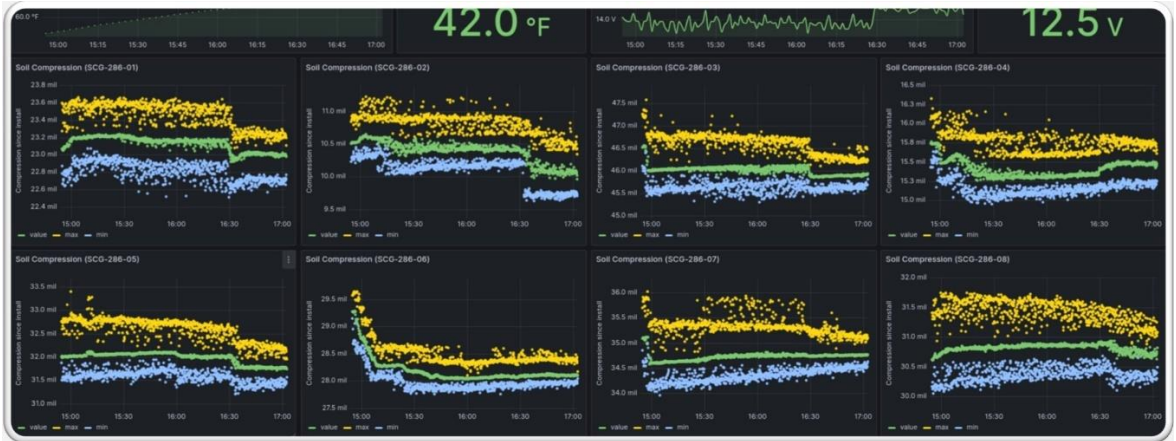

**Figure 19 A sample Grafana dashboard streaming live sensor data**

Figure 20 provides a comprehensive overview of the end-to-end system, highlighting the interactions between the individual modules involved in the I-65 field experiments. This illustrates the practical realization of a specific solution within the overarching real-time data management architecture and outlines the system workflow to facilitate the operation of the validated instance.

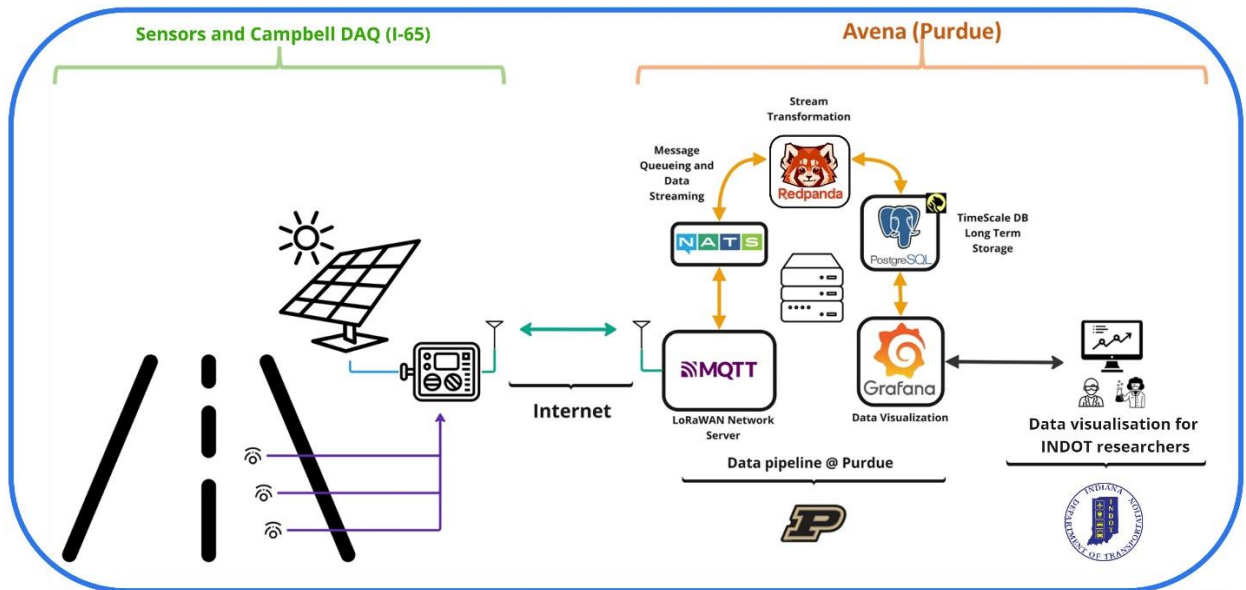

**Figure 20 End-to-end connectivity of system components in the I-65 field trial**

## RESULTS

It was observed that the real-time system architecture seamlessly integrates to enable a robust, real-time data processing system for the I-65 field trials. The case study system built on a sophisticated framework like Avena achieved low latency sensor data exchange with the server. Open-source tools like TimescaleDB supported live data storage and Grafana enabled impactful data visualization and real-time sensor monitoring.

Although the static dataset management solution describes generic guidelines to manage local databases, it offers great flexibility in project data management, as evident from the diverse nature of the projects included under its scope in the case study. The use of relational databases like Oracle guarantees ACID (atomicity, consistency, isolation, and durability) properties while managing the critical synopsis of the large database. Blob storage solutions enable economical archiving of gigabytes of unprocessed sensor data at no or low costs. Specialized data visualization tools like ArcGIS addressed the project specific data visualization demands of the researchers.

It is essential to highlight that the actual insights derived from the processed data are of limited importance for the current work, but rather the key aspect of the case studies is to demonstrate how the solution fits into the proposed system guidelines.

## CONCLUSIONS

In this work we conducted a deep dive analysis of the best practices for the effective management of sensor data collected from field testing of the instrumented sections of I-65 and I-69 Greenfield districts. The data was collected by Purdue-INDOT researchers to monitor and evaluate pavement behavior over time. In the study, we proposed that for massive historical datasets, summary and metric data are suitable for storage in relational databases and the unprocessed data was more suited to blob storage. Researchers primarily prefer sensor metrics over the raw sensor data itself. They reduce the data complexity by applying the analytical methods of prior research to the data before storing it in relational databases like Oracle and utilizing it within existing INDOT data systems. Such systems include FME and ArcGIS Online.

Finally, a complete end-to-end system architecture for collecting embedded sensor data in real-time via remote data acquisition systems was proposed. The case study system used solar-powered Campbell equipment which is largely designed for automatic but local data collection. The data management solution leveraged the Avena framework which modernizes the design of real-time remote data collection and enables edge computing on the data stream. As a result, the sensors can be monitored 24x7, and useful metrics produced in real-time and pushed to centralized data storage, massively reducing the latency between measurement and use in INDOT data-driven decisions. However, the restrictions of the Campbell DAQ hardware motivate further exploration into potentialities of introducing Avena within the roadside DAQ to enable a more open data flow.

The system proposals provide holistic end-to-end solutions for static and real-time data management in transportation research studies. The presented case studies validate the functionality and the effectiveness of the proposals and fosters confidence in its operation. The work will encourage the researchers to adopt and integrate the proposals into various real-world projects within and across (13-14) the boundaries of transportation research.

## ACKNOWLEDGMENTS

This work was supported in part by the Joint Transportation Research Program administered by the Indiana Department of Transportation and Purdue University. The contents of this paper reflect the views of the authors, who are responsible for the facts and the accuracy of the data presented herein. The contents do not necessarily reflect the official views and policies of the Indiana Department of Transportation or the Federal Highway Administration. The paper does not constitute a standard, specification or regulation.

## AUTHOR CONTRIBUTIONS

The authors confirm contribution to the paper as follows: study conception: Shin Boonam, James. V. Krogmeier; system design and workflow planning : Andrew Balmos; analysis, interpretation of results and draft manuscript preparation : Manish Kumar. All authors reviewed the results and approved the final version of the manuscript.